# Mechanical Evidence for the Phylogenetic Origin of the Red Panda's False Thumb as an Adaptation to Arboreal Locomotion

Braden Barnett, Yiqi Lyu, Kyle Pichney, Brian Sun, Jixiao Wu

## I. Abstract

We constructed a modular, biomimetic red panda paw with which to experimentally investigate the evolutionary reason for the existence of the false thumbs of red pandas. These thumbs were once believed to have shared a common origin with the similar false thumbs of giant pandas; however, the discovery of a carnivorous fossil ancestor of the red panda that had false thumbs implies that the red panda did not evolve its thumbs to assist in eating bamboo, as the giant panda did, but rather evolved its thumbs for some other purpose. The leading proposal for this purpose is that the thumbs developed to aid arboreal locomotion. To test this hypothesis, we conducted grasp tests on rods 5-15 mm in diameter using a biomimetic paw with 0-16 mm interchangeable thumb lengths. The results of these tests demonstrated an optimal thumb length of 7 mm, which is just above that of the red panda's true thumb length of 5.5 mm. Given trends in the data that suggest that smaller thumbs are better suited to grasping larger diameter rods, we conclude that the red panda's thumb being sized below the optimum length suggests an adaptation toward grasping branches as opposed to relatively thinner food items, supporting the new proposal that the red panda's thumbs are an adaptation primary to climbing rather than food manipulation.

## II. Introduction

We sought to clarify the evolutionary history behind the development of the unique false thumbs of modern red pandas to determine if these structures truly evolved independently of the similar structures seen in the giant panda.

This question is of interest to the field of phylogenetics, the study of evolutionary history, because if the false thumbs of the two pandas are not related, then the red panda and giant panda represent a remarkable example of convergent evolution. If true, this would place the only two known panda species evolutionarily distant from each other, as depicted in recent evolutionary trees such as that in Fig. 1A [1].

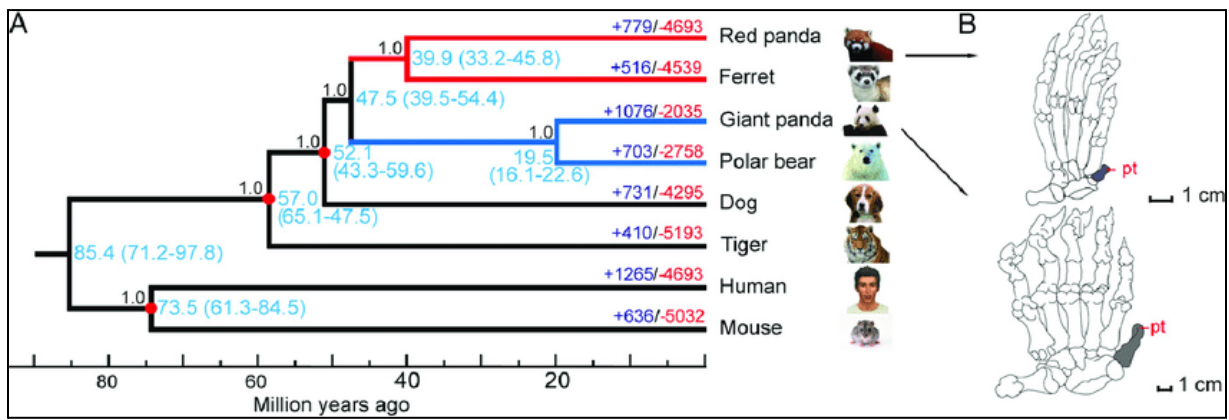

Fig. 1. (A) A proposed evolutionary tree with the red panda and giant panda not closely related. (B) The skeletal structure of the paw of the red panda (top) and giant panda (bottom) with their radial-sesamoid-derived false thumbs shaded. Reproduced from [1].

The evolutionary tree in the figure, however, is just a recent one in a line of many contradictory trees that have been proposed by phylogeneticists attempting to understand the relatedness of the two pandas. Much of the contention originates from the structural and behavioral similarities of the two pandas [2].

Structurally, it can be seen in Fig 1B that the false thumbs of both pandas are similar. In both pandas, the false thumbs exist in addition to a five-fingered hand and are not true digits, but rather extensions of a wrist bone, specifically, the radial sesamoid [1].

Behaviorally, both pandas are herbivorous, and both have been observed using their false thumbs to manipulate bamboo, a staple of their diets. From these observations, it is only natural that early attempts to classify the giant panda and red panda placed them together in the ursidae family [2].

This theory of a common origin; however, was called into question when a fossil ancestor of the red panda, depicted in Fig 2., was discovered with false thumbs. This ancestor was carnivorous,and therefore implies that the red panda developed its false thumbs before it would have needed them to manipulate and eat bamboo. Based on this discovery, and the fact that both the red panda and its ancestor are arboreal, the red panda's false thumbs have been reinterpreted as an adaptation to climbing thin branches rather than one for food manipulation [3].

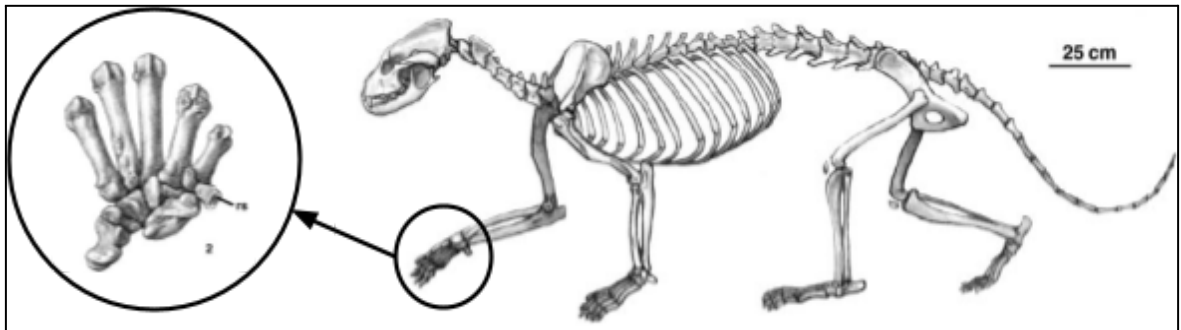

Fig. 2. S. batelleri, a carnivorous and arboreal fossil ancestor of the red panda which has a radial-sesamoid-derived false thumb. Adapted from [3].

This interpretation, however, like many evolutionary theories, has not been quantitatively tested, but is based on the principle of parsimony analysis. Parsimony analysis is the standard method of determining evolutionary relationships within the field of phylogenetics. It consists of building a list of characteristics, either physical or genetic, across various species, and determining the likeliest evolutionary trajectory that would have required the fewest changes to branch off each species from a common ancestor. While this technique is a powerful one, it does not answer the question of why a change would have occurred. That is, it says nothing about the nature of the selection pressure driving evolution. To make a statement on why the red panda developed its false thumbs, it is necessary to evaluate the nature of this selection pressure.

This project seeks to perform the quantitative testing that is absent from parsimony analysis. It serves as an example of how biomimetic robotics can be used to experimentally evaluate evolutionary theories, particularly on extinct or endangered animals for which biological testing is difficult or impossible. By replicating biological structures with a mechanical facsimile, modular designs can be used to recreate a proposed evolutionary trajectory. This enables experimental data to be gathered by which to quantify the marginal benefits of evolutionary changes.

In this project, we replicate the evolutionary trajectory of the red panda's false thumbs. We do this through the design of a modular and biomimetic panda paw with variable length thumbs. As we progressively lengthen the thumbs of our mechanical panda paw, we evaluate the incremental benefit it offers to the red panda's ability to grasp objects of various diameters.

Our experiments clarify the relationship between thumb size and favored grip size. By observation of this relationship, we are able to determine if the red panda's thumb favors gripping branch-sized items or food-sized items. In doing this, we determine if the benefits offered by the red panda's thumb reflect an evolutionary purpose of serving as an aid to climbing, as now proposed, or an aid to food manipulation, as believed for the giant panda.

### III. Research Question

Our investigation was driven by a central research question:

Does the red panda's false thumb offer sufficient benefit to the animal's ability to climb thin branches that natural selection would have driven the thumb's lengthening from the radial sesamoid bone for this reason alone?

To evaluate this question, we pose the hypothesis:

As the red panda's thumb is lengthened, the minimum diameter object that can be securely grasped will decrease, reflecting an enhanced ability to grasp thin branches.

### IV. Design

To create a mechanical paw representative of the true paw, we first had to understand the red panda's biology. Toward this end, the team sought out anatomical guides to the red panda and found detailed depictions of the red panda's skeletal structure.

As we wanted to make our design as realistic as possible, we chose to design our paw to scale. We found that the length of a red panda from snout to base of the tail ranged from 510 to 635 mm [4]. Selecting the high side of this range to match later conservative estimates of panda weight, we estimated the dimensions of the red panda's forelimb, and from that, the dimensions of each phalanx of the paw, as shown in Fig. 3.

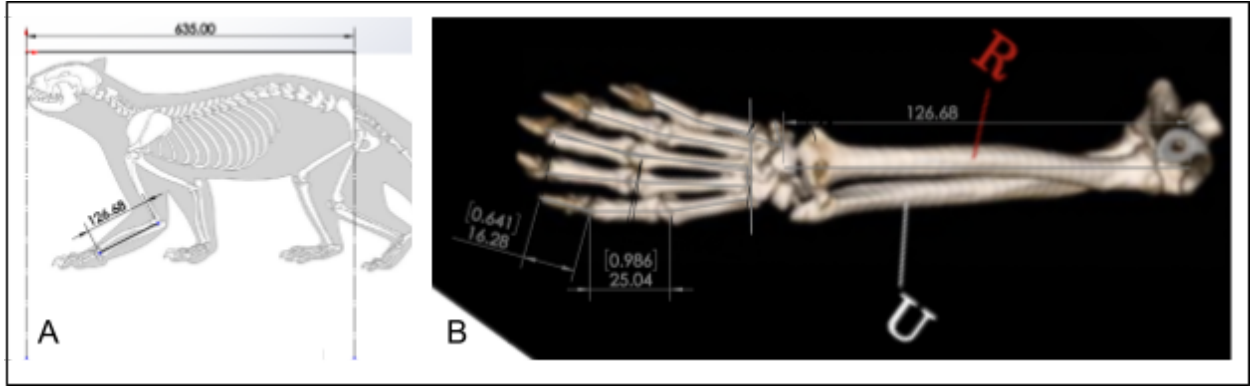

*Fig. 3. Dimensional approximation. (A) Arm length approximation from body length. Adapted from [4]. (B) Phalanx approximation from arm length. Adapted from [5].*

These dimensions were replicated in the design of a 3D-printed mechanical hand. Each of the five phalanges of this mechanical hand were designed to consist of jointed segments reflective of the structure of the true paw. PETG was chosen as the print material due to its increased strength, stiffness and fatigue life over PLA. To replicate the tendons that drive the real hand, we chose to actuate the fingers of our mechanical paw using a cable drive.

To determine the force our mechanical paw would need to output, we analyzed the grip strength of living red pandas. As this data was not readily available, we observed the locomotion of red pandas in their arboreal environments and noted that, at any given time, at least two of the panda's four limbs were being used to support the animal on a branch. Based on this observation, we assumed that our paw would need to output sufficient grip force to support a load equal to half of the red panda's total body weight.

For a conservative estimate of grip force, we used a red panda body mass of 6.2 kg, the greatest reported in [4]. Assuming a uniform pressure distribution of the force exerted by the hand on a vertical rod in its grasp, the free body diagram in Fig. 4 was generated.

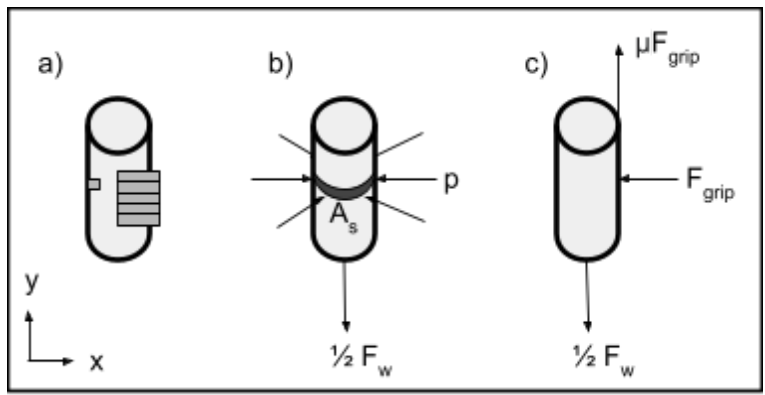

*Fig. 4. Estimation of the required grip force assuming a uniform pressure distribution and a maximum supportable weight of ½ of the red panda's total weight.*

From this diagram, we estimated that the grip force must be equal to half of the red panda's body weight divided by the coefficient of static friction between the rod and the hand. As this friction coefficient was not readily available, we again made a conservative estimate, taking the coefficient of friction as 0.3, the lowest value reported for leather on wood [6]. This yielded a required grip force of 101.4 N.

To simulate our assumed friction coefficient and better approximate the compliance of the flesh and skin of the real animal, we covered all contact points of our mechanical paw with Dragonskin 30, a soft material with skin-like properties.

The final challenge of the design was understanding the mechanics of the false thumb. On this point, the available biological research is contradictory. One interpretation proposes that the false thumb merely acts as a supporting ridge with which to grasp objects. Another suggests that the false thumb and its musculature, depicted in Fig. 5, are involved with the supination of the forearm and adduction of the palm [7]. As the biology is uncertain, we elected to follow the first interpretation, which implies a simpler mechanism, and left the second interpretation as a stretch goal for future work.

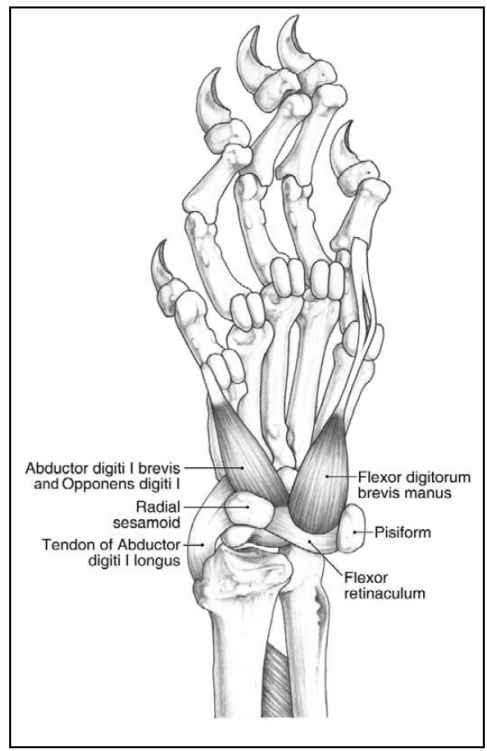

*Fig. 5. Skeletal and muscular structure of the red panda's paw. Reproduced from [4].*

With the first interpretation selected, we designed an interface for interchangeable thumbs on the side of the palm. Once a thumb was attached to this interface, it could be pivoted into or out of place. This enabled testing with and without the thumb acting as a supporting ridge.

The final biomimetic paw and test apparatus design are pictured in Fig. 6 with major design features called out. The test structure was designed from 80/20 aluminum extrusions. The cable drive was implemented by running fishing line through channels built into each phalanx and running each line to a cylindrical spool turned by a stepper motor. The stepper motor was selected to be capable of achieving the 101.4 N grip strength derived earlier while also being able to move in high resolution, discrete steps to simplify control.

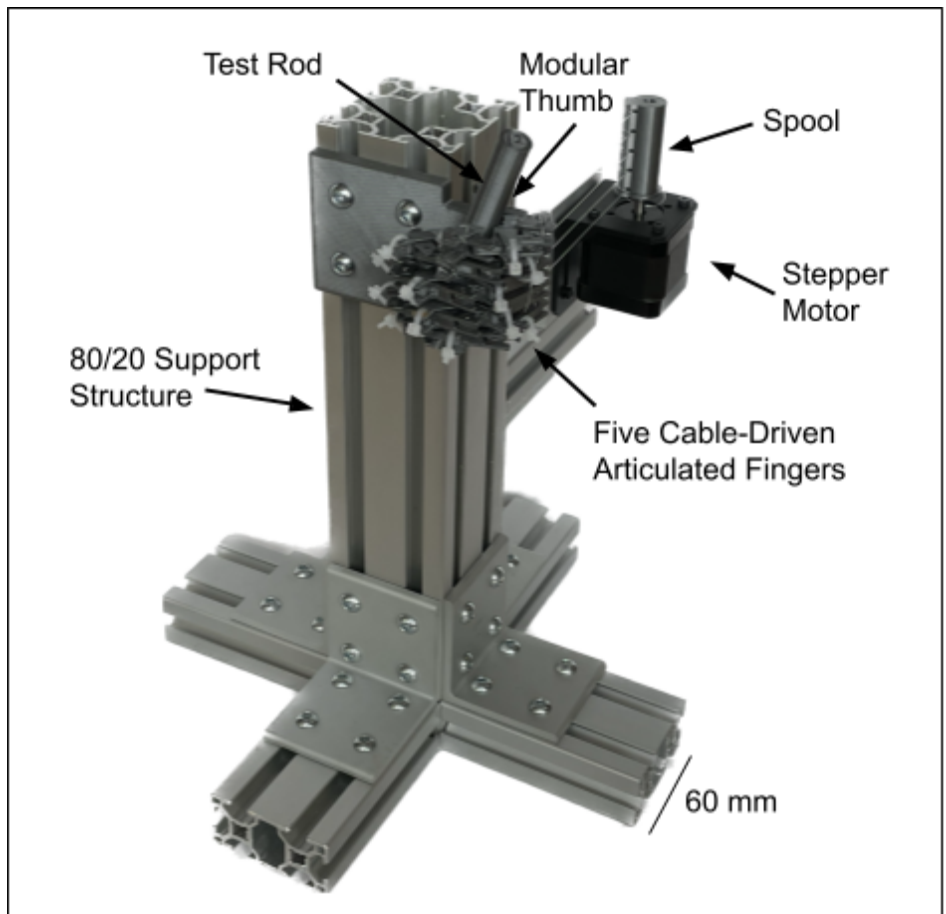

Fig. 6. The final test apparatus and biomimetic paw design in a representative grasp test configuration. The paw consists of five wire-driven articulated fingers with a modular false thumb. Actuation is enabled via a stepper motor.

Although this design did not directly achieve the stretch goals from the proposal, it did satisfactorily achieve all of the primary design objectives laid out at the start of the project. Furthermore, with additional time, several of the stretch goals could be readily met. For instance, with our modular thumb design, we could easily compare the thumb designs of the red panda, giant panda, and fossil ancestor after obtaining accurate estimates of their dimensions. Additionally, we completed all of the electrical and code set up for grasp force testing and would only need additional time to finish working out the physical set up.

## V. Methods

To evaluate the effect of thumb length on the panda's ability to grasp objects, we designed a grasp test using eleven test rods and five modular thumbs, as depicted in Fig. 7. The biomimetic paw attempted to grasp each rod without a thumb and with each thumb design. The test procedure is as described below.

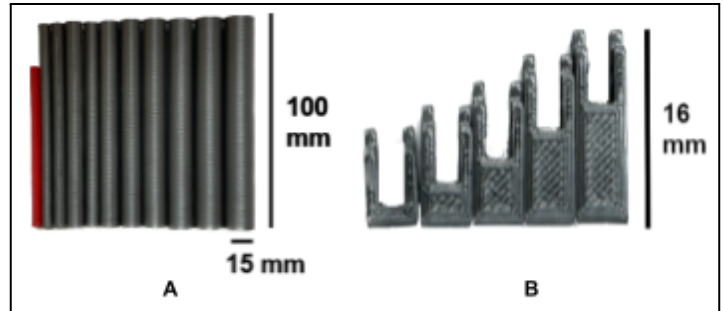

*Fig. 7. (A) The eleven grasp test rods ranging in diameter from 5 to 15 mm. (B) The five test thumbs ranging in length from 5.5 to 16 mm. Note: Fig. 7A and 6B are not depicted to scale with each other.*

We start with the robot hand open and hold the rod of the diameter we are testing against the palm between the thumb and fingers. We place the rod such that it is touching the tip of the thumb, or in the case of no thumb, directly against the palm in the location where the thumb would touch. We then actuate the paw closed in 200 step increments using the GUI controls pictured in supplemental figure A2 in the Appendix. We keep closing the paw until the motor slips. We then actuate by 100 steps until it slips again. We work our way down to 5 step increments until the grip is tight. We now evaluate two criteria: the quality of the contact between the thumb and rod; and how well the rod is being held. We had three definitions for contact:

- Good contact: The rod is held between the tip of the thumb and at least one finger
- Bad contact: The rod is held but with contact on the side of the thumb
- No contact: The rod is held but with no contact with the thumb

To evaluate the strength of the grip, we tapped on the end of the rod to see if it would fall out or not. If the rod had minimal movement such that it could be dexterously manipulated by a real hand, we considered the grasp to be held. If the rod would slide easily but did not fall out, we considered the grasp to be loosely held. If the rod fell out completely before being tapped, the rod was not held. We would then release the grip and perform another trial using the same step size. We attempted 3 trials for each rod.

Because our data was largely qualitative in nature, it was difficult to perform traditional statistical analyses. Nevertheless, with three trials of each thumb and rod combination, we were able to demonstrate consistency in our test results. Three trials were selected on account of the number of thumb-length/rod-diameter combinations we tested. This is a traditional minimum number of trials. With more time, however, we would increase this number to resolve the few apparent outliers we did observe.

## VI. Results

The results of our grasp tests are presented in Table 1. Cells with a green coloration represent successful grasps that were well-assisted by the thumb. Cells with a beige or yellow coloration represent grasps that held the rod but were not well assisted by the presence of the thumb. Cells with a red coloration represent rod and thumb combinations for which the rod could not be held by the paw.

*Table 1. Grasp Test Results for Three Trials of Each Thumb/Rod Combination.*

Across the three trials for each thumb/rod combination, we see notable outliers only in two trials: the 8 mm rod with no thumb and the 11 mm rod with no thumb. With only three trials for each combination and a dataset that is largely qualitative in nature, statistical analysis is difficult for these results. Nevertheless, the consistency of our results both across trials and between experimental neighbors makes us confident in our results. With more experimentation, we believe we would see the same results.

Analyzing the data, we can conclude that the optimal thumb size for our experiment was 7 mm. With this thumb, all rod diameters could be grasped successfully with the thumb well engaged. We also observe an inverse relationship between the length of the thumb and the rod diameters the paw can best grip. For small thumb lengths, the paw was better able to grip larger diameter rods. For large thumb lengths, the paw was better able to grip smaller diameter rods.

Together, these results support our hypothesis. As thumb length increased, smaller diameter rods could be grasped. Our data also support the theory of the red panda's thumb being an adaptation to climbing rather than food manipulation. Because the red panda's thumb is sized near to but slightly below the optimum, the thumb exhibits a preference toward grasping larger diameter objects, such as branches, as opposed to the smaller diameter stalks and bamboo shoots it manipulates for food.

## VII. Conclusions & Future Work

Our project aimed to investigate whether longer thumbs would enable a robot, based on a red panda, to grasp thinner objects. Our findings indicate that this is indeed the case, as the optimal thumb length for our setup closely matched that of red pandas. However, we also discovered an unexpected outcome: longer thumb lengths impeded the robot's ability to hold larger diameter rods. This could explain why the red panda does not have longer thumbs more in line with those of the giant panda. The current length of the red panda's thumbs allows them to grasp a wider range of diameters, in particular, those diameters that are representative of the branches they climb along. This result supports the theory that the red panda's thumb is primarily adapted for arboreal locomotion, and only secondarily applied to food manipulation.

Future work on this project would focus on reaching the stretch goals we had initially laid out and then continuing to enhance the realism of our design to draw better conclusions on the nature of the red panda's thumb. First, we would like to get the force sensing to work properly. The sensor is just slightly too large for our paw to hold as is. The easiest way we imagine resolving this would be to manufacture some form of plate with an offset level. With this, the paw could grip onto a narrower stage and transfer its force to a wider stage around the sensor. Second, we would like to make general improvements to the realism of the design. Ideally, we would get accurate measurements of the properties of the red panda's skin and fur and manufacture a near-identical replica. In addition, we would also change from our stepper motor and cable driven actuation method to soft actuators to mimic real muscle. Finally, as the biological mechanism of how the red panda uses its false thumb remains contentious, we would also be interested in weighing in on this debate by designing false thumbs that act just as a ridge, as we have done in these experiments, and comparing them to a design in which the thumb is powered with an active range of motion. By testing both, we could help to resolve this ambiguity by exploring the mechanical feasibility of both options given the biological structures seen in the animal.

## VIII. Appendix

### A. Supplemental figures

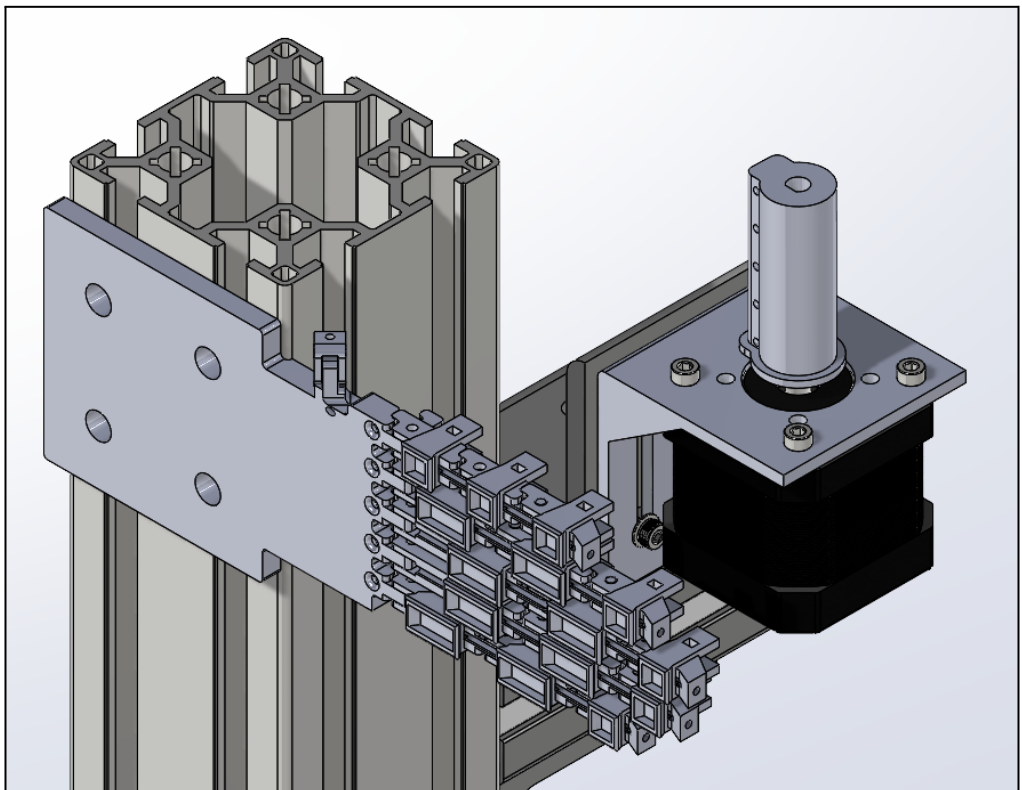

*Fig. A1. Final CAD Model in SolidWorks.*

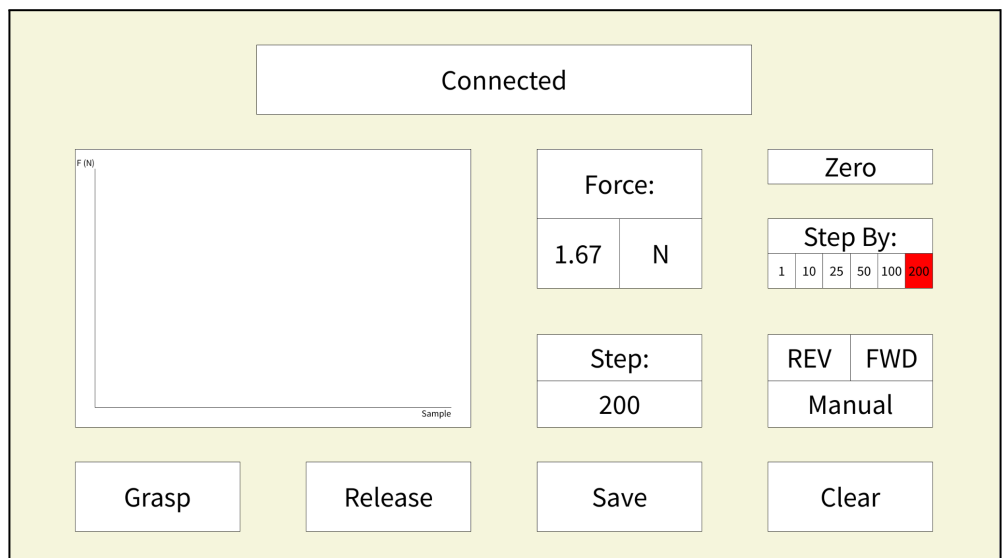

*Fig. A2. Final GUI used to interface with the system.*

Grasp Control and Interface Pseudocode

| Grasp Control (Arduino) | Interface (Processing) |
|---|---|
| 1. Initialize all input/output pins<br>2. Zero the force sensor<br>3. Read serial data and if the data is:<br>   a. "Connect" → Confirm connection<br>   b. "Grasp" → Close hand until grip force exceeds a threshold<br>   c. "Release" → Return the actuator to the zero position<br>   d. "Manual" → Step forward/reverse according to the encoder value and the set step increment<br>   e. "Zero" → Take 1000 force readings and set an offset to zero the average<br>   f. Numeric → Step forward/reverse by the specified number of steps<br>   g. Otherwise → Print the force and step measurements once per second | 1. Define all visual elements<br>2. Check if the current program state should be changed<br>3. Check buttons and if the button clicked is:<br>   a. "Grasp" → Write to serial: "Grasp"<br>   b. "Release" → Write to serial: "Release"<br>   c. "Manual" → Write to serial: "Manual"<br>   d. "Zero" → Write to serial: "Zero"<br>   e. Numeric → Update the step increment<br>   f. "FWD" → Write to serial: the step increment as a positive number<br>   g. "REV" → Write to serial: the step increment as a negative number<br>   h. "Save" → Write stored data to a CSV<br>   i. "Clear" → Empy stored data<br>4. Read serial data; update visual elements, the step count, and the force reading |

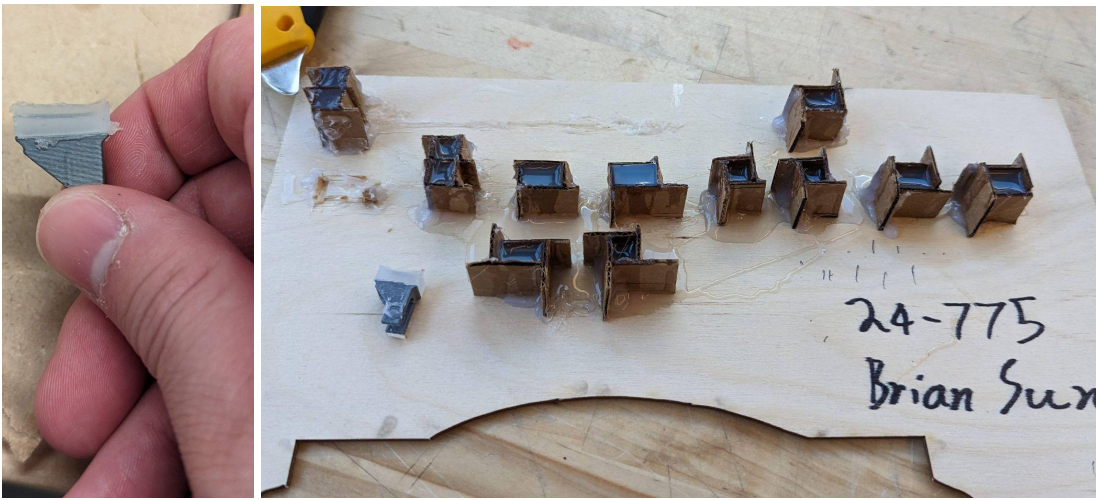

*Fig. A3: Soft material pad fabrication.*

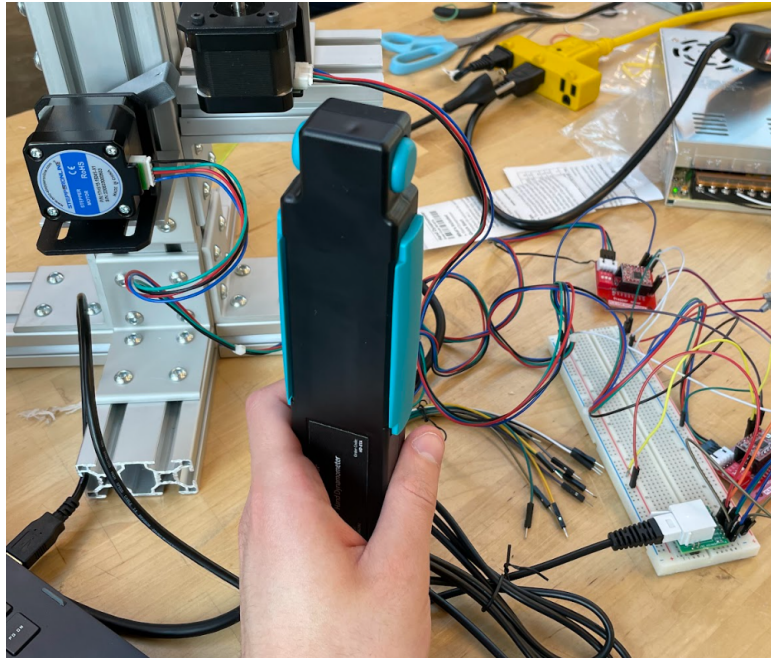

*Fig. A4: Force sensing apparatus.*

## B. Bill of Materials

The team's final project BOM is presented in Table B1. Our largest expenses were the Vernier Hand Dynamometer, which we selected to ensure that we had a high quality grip force sensor, and the 80/20 structural components, which we used to mount our paw and actuator within a stable test apparatus.

*Table B1. Project BOM.*

| Subsystem | Name | Quantity |
|---|---|---|
| Skin | Dragon Skin 30A | |
| Paw Mechanism | PETG Filament(2 spools) | 2 |
| Control | A4988 Motor drivers | 1 |
| Control | Control Extension Shield | 1 |
| Control | Arduino Uno | 1 |
| Control | Stepper Motor 17HS15-1504S | 2 |
| Test apparatus | Hand Dynamometer | 1 |
| Test apparatus | Analog Protoboard Adapter | 1 |
| Support | T-Slotted Framing Silver Corner Bracket, 2-3/8" Long, 60 mm Double/Quad Rail | 7 |
| Support | T-Slotted Framing Quad Rail, Silver, 60 mm High x 60 mm Wide, Hollow, 4' Long | 1 |
| Support | T-Slotted Framing Double 6-Slot Rail, Silver, 60mm High, 30mm Wide, Hollow 3' Long | 1 |
| Support | Cable Tie Narrow, 3" Long, 18 lbs. Breaking Strength, Off-White | 1 |
| Support | Alloy Steel Socket Head Screw Black-Oxide, M3 x 0.5 mm Thread, 6 mm Long | 1 |
| Support | Black-Oxide Steel Hex Nut Medium-Strength, Class 8, M3 x 0.5 mm Thread | 1 |
| Support | Black-Oxide 18-8 Stainless Steel Washer for M3 Screw Size, 3.2 mm ID, 7 mm OD | 1 |
| Support | Alloy Steel Socket Head Screw Black-Oxide, M3 x 0.5 mm Thread, 15 mm Long | 1 |
| Support | Black-Oxide Steel Oversized Washer for M3 Screw Size, 3.2 mm ID, 12 mm OD | 2 |
| Support | Mounting Bracket for Nema 17 Stepper Motor | 2 |
| Control | 24V DC Power Supply | 1 |
| Paw Mechanism | 18-8 Stainless Steel Slotted Spring Pin 2mm Diameter, 6mm Long | 2 |
| Paw Mechanism | Spring-Back Multipurpose 304 Stainless Steel Wire 1/4 lb. Coil, Matte, 0.039" Diameter | 1 |
| Paw Mechanism | uxcell 1mm Wire Rope Aluminum Sleeves Clip Fittings Cable Crimps 100pcs | 1 |
| Paw Mechanism | Dowel Pin, 4140 Alloy Steel, 2 mm Diameter, 45 mm Long | 2 |
| Power Supply | Power Cord | 1 |

## C. References